\documentstyle[12pt]{article}
\begin{document}
\def\om{\omega}
\def\omt{\tilde{\omega}}
\def\ti{\tilde}
\def\o{\Omega}
\def\t{T^*M}
\def\vt{\tilde{v}}
\def\ot{\tilde{\Omega}}
\def\otwo{\omt \wedge \om}
\def\owot{\om \wedge \omt}
\def\w{\wedge}
\def\mt{\tilde{M}}
\def\ss{\subset}
\def\tpm{T_{P} ^* M}
\def\al{\alpha}
\def\alt{\tilde{\alpha}}
\def\la{\langle}
\def\ra{\rangle}
\def\inop{{\int}^{P}_{P_{0}}{\om}}
\def\th{\theta}
\def\tht{\tilde{\theta}}
\def\inox{{\int}^{X}{\om}}
\def\inotx{{\int}^{X}{\omt}}
\def\st{\tilde{S}}
\def\l{\lambda}
\def\p{{\bf{p}}}
\def\pb{{\p}_{b}(t,u)}
\def\pbm{{\p}_{b}}
\def\d{\partial}
\def\d+{\partial_+}
\def\d-{\partial_-}
\def\dpm{\partial_{\pm}}
\def\l2{\Lambda^2}
\def\be{\begin{equation}}
\def\ee{\end{equation}}
\def\ej{{\bf E}}
\def\ed{{\bf E}^\perp}
\def\si{\sigma}
\def\cg{{\cal G}}
\def\cgt{\ti{\cal G}}
\def\cd{{\cal D}}
\def\ce{{\cal E}}
\def\bz{\bar{z}}
\def\e{\varepsilon}
\def\b{\beta}
\begin{titlepage}
\begin{flushright}
CERN-TH/95-251\\
ETH-TH/95-26\\
Imperial/TP/94-95/61\\
UUITP-16/95  \\
hep-th/9509123
\end{flushright}
\begin{center}
{\Large\bf Quantum Poisson-Lie T-duality}\\
{\Large\bf and  WZNW model} \\
\vskip 1cm
{\bf A. Yu. Alekseev\footnote{On leave of absence from Steklov
Institute, St.-Petersburg}}\\
\vskip 0.2cm
{\it Institute for Theoretical Physics, Uppsala University, Box 803}\\
{\it S-75108, Uppsala, Sweden\footnote{Permanent address}}\\
\vskip 0.1cm
{\it and}
\vskip 0.1cm
{\it Institut f\"ur Theoretische Physik, ETH, CH-8093 Z\"urich, Switzerland}\\
\vskip 0.3cm
{\bf C. Klim\v c\'\i k} \\
\vskip 0.2cm
  {\it Theory Division CERN, CH-1211 Geneva 23,
Switzerland} \\
\vskip 0.2cm
and\\
\vskip 0.2cm
{\bf A. A. Tseytlin\footnote{On leave of absence from Lebedev Institute,
Moscow}}\\
\vskip 0.2cm
{\it Theoretical Physics Group, Blackett Laboratory, Imperial College,}\\
{\it London SW7 2BZ, UK}
\end{center}

\begin{abstract}

A pair of conformal $\si$-models related by Poisson-Lie T-duality is
constructed by starting with the
 $O(2,2)$ Drinfeld double. The duality relates the standard
$SL(2,R)$ WZNW model to a constrained $\sigma$-model defined on the  $SL(2,R)$
group space.
The quantum equivalence of the models is established
by using a path integral argument.

\end{abstract}
\noindent CERN-TH/95-251\\
September 1995
\end{titlepage}

\section{\bf Introduction}

Target space duality in string theory has attracted a considerable
attention in recent years because it sheds some  light on the
geometry and symmetries  of string theory.
Much is known about the standard Abelian  $\sigma$-model T-duality
\cite{Busch,GiRa,Ts,RoVe,Alv1,GPoR,AAL2}.
 However,  the structure and role
   of the non-Abelian duality  still remains
to be uncovered.  The non-Abelian duality
between the isometric $\si$-model on a group manifold $G$ and the non-isometric
$\si$-model on its Lie algebra $\cg$ discussed in
\cite{FJ,FT,OQ,GR}
 did miss a lot of features characteristic  to
 the Abelian duality.
One  could  even question if the
term  `duality'  was  applicable since
the original and the dual $\si$-models
did  not  enter  the picture symmetrically.
 Indeed, while the original
model on $G$ was isometric, which was believed to be  an essential condition
for performing a  duality transformation, the `dual' one  did not possess the
$G$-isometry. As a result,   it was not known how to perform the  inverse
duality transformation
to get back to the original model.

A solution to this  problem  was proposed recently in \cite{KS2} where it was
argued that the two theories are, indeed,  {\it dual} to each other
from the point of view of  the so called Poisson-Lie T-duality  (this term was
introduced later in \cite{K}). In \cite{KS2} a large
class of new dual pairs of $\si$-models  associated with
each Drinfeld double $D$ \cite{D} (or, more precisely, with
 each Manin triple
$(\cg,\cgt,\cd)$
corresponding to $D$) was constructed.
 The main idea of the approach is to replace
the so far  key feature of the T-duality - the requirement
of isometry -   by a weaker condition  which is  the Poisson-Lie
symmetry of the  theory. This generalized
 duality is  associated with the  two groups forming
a Drinfeld double and the duality transformation  exchanges their  roles.
We shall review  some  elements of the Poisson-Lie T-duality  in  section 2.

The discussion in \cite{KS2} was purely classical. It is obviously of central
importance to try to understand if there is a quantum analogue of the
Poisson-Lie T-duality relating  appropriate  correlation functions in the two
models.
In particular, one would like to know if
 there are  dual pairs of
conformal $\si$-models.  This  is the question we address in the present paper.
As we shall show in section 3,
there exists  a simple  example  of such dual pair  associated with  the
 $O(2,2)$ Drinfeld double. Here
$G=SL(2,R)$ and the corresponding  model is a constrained
$\si$-model with a target space being $SL(2,R)$ group manifold. Its dual
associated with
$\ti G =B_2$  (where $B_2$ is the Borel subgroup of $SL(2,C)$)  is the
$SL(2,R)$ WZNW model.
In the section 4   we  shall give  a path integral argument demonstrating
 quantum
equivalence of the two  such  dual models
 related to the groups $G$ and
$\ti G$.
A detailed account of our approach and its range of applicability
 will appear later.

\section{\bf Poisson-Lie T-duality}

In this section  we  will describe the construction of a dual pair
of $\si$-models which  have  equivalent  field equations (and  symplectic
structure of their
phase spaces \cite{KS2})   in the sense  that there exists a well defined
 (though possibly nonlocal) transformation which  to  every
solution of the first model associates a solution of the dual one
and vice versa.
They are dual in a new `Poisson-Lie' sense
which generalizes the
Abelian T-duality \cite{Busch} and  the non-Abelian T-duality between
$\si$-models on a group and  on its Lie algebra
\cite{FJ,FT,OQ,GR}.\footnote{In \cite{Ty} there was suggested  a
connection  between
non-Abelian axial-vector duality \cite{KO} and Poisson-Lie T-duality.}
For simplicity we  shall consider only the case of the $\si$-models defined on
the group space; generalization  to the case
 when a group  $G$ acts freely
on the target space ($G$-bundle),  the Abelian analogue of which was
 discussed  in \cite{Busch,RoVe}, was given in  \cite{KS2,K}.

 For the description of the Poisson-Lie duality we need the
crucial concept
of the Drinfeld double  which is simply  a  Lie group $D$ such that
its Lie algebra
$\cd$ can be decomposed into a pair of maximally isotropic subalgebras with
respect to a non-degenerate invariant bilinear form on $\cd$ \cite{D}.
 An isotropic subspace of $\cd$ is such that the value of the invariant form
on any
two vectors belonging to the subspace vanishes (maximally isotropic means
that this subspace cannot be enlarged  while preserving its  isotropy).
 Any such decomposition of the double into
the pair of maximally
isotropic subalgebras
$\cg + \cgt=\cd $ is usually referred to as the Manin triple.

Consider now an $n$-dimensional linear subspace $\ce^+$ of the Lie algebra
$\cd$ and its orthogonal complement $\ce^-$ such that
$\ce^+ + \ce^-$ span the whole algebra $\cd $. We shall show that these
data determine a dual pair of  $\si$-models with the targets being
the groups $G$ and $\ti G$ respectively \cite{KS2,K}. Indeed, consider
the following {\it field equations}
 for the mapping $l(\xi^+,\xi^-)$ from the
two-dimensional spacetime with light-cone variables $\xi^{\pm}$
into the Drinfeld double group $D$
\be \langle \dpm l \ l^{-1},\ce^{\mp}\rangle=0  . \ee
Here the brackets denote the invariant bilinear form.
In the vicinity of the unit
element of $D$, there exists
the unique decomposition of an arbitrary element of $D$ into
 the product of elements
from $G$ and $\ti G$, i.e.
\be l(\xi^+,\xi^-)=g(\xi^+,\xi^-)\ti h(\xi^+,\xi^-).\ee
Inserting this ansatz into Eq. (1) we obtain
\be \langle g^{-1}\dpm g +\dpm \ti h \ti h^{-1},g^{-1}\ce^{\mp}g\rangle=0.\ee
It is convenient to introduce a pair of  bases
$T^i$ and $\ti T_i$ in the algebras $\cg$ and
$\cgt$ respectively, satisfying the duality condition
\be \langle T^i,\ti T_j\rangle =\delta^i_j.\ee
Suppose there exists a matrix $E^{ij}(g)$ such  that ($i,j=1,\dots,n$)
\be g^{-1}\ce^+ g ={\rm Span}(T^i + E^{ij}(g)\ti T_j) ,\ee
\be g^{-1}\ce^- g ={\rm Span}(T^i - E^{ji}(g)\ti T_j).\ee
The explicit dependence of the matrix $E$ on $g$ is given
by the matrices of the adjoint representation of $D$ and  is  easily
obtained
 as follows
  \cite{KS2,K}:
\begin{eqnarray} g^{-1}\ce^+ g={\rm Span}~ g^{-1}(T^i + E^{ij}(e)\ti T_j)g\cr
={\rm Span}\Big[(a(g)^i_{~l} +E^{ij}(e)b(g)_{jl})T^l +
E^{ij}(e)d(g)_j^{~l} \ti T_l\Big],\end{eqnarray}
where
\be g^{-1}T^i g\equiv a(g)^i_{~l} T^l, \qquad g^{-1}\ti T_j g \equiv
b(g)_{jl}T^l +d(g)_j^{~l} \ti T_l.\ee
Hence the  matrix $E(g)$ is given by
\be E(g)=[a(g) + E(e)b(g)]^{-1}E(e)d(g).\ee
We can now rewrite the field equations (3) in the form
\be -(\partial_-\ti h \ti h^{-1})^i=
E^{ij}(g)(g^{-1}\partial_- g)_j\equiv A_-^i(g),\ee
\be -(\partial_+\ti h \ti h^{-1})^i=
-E^{ji}(g)(g^{-1}\partial_+ g)_j\equiv A_+^i(g).\ee
This implies the  `zero curvature' condition for $A^i(g)$:
\be \partial_- A_+^i(g) -\partial_+ A_-^i(g) -
\ti c_{kl}^{~~~i} A_+^k(g) A_-^l(g)=0,\ee
where $\ti c_{kl}^{~~~i}$ are the structure constants of the Lie algebra
$\cgt$. Remarkably,  it can be checked  directly that the  equations  (12) are
just the field equations for  the $\si$-model on the group space $G$
with the Lagrangian
\be L=E^{ij}(g)(g^{-1}\partial_+ g)_i(g^{-1}\partial_- g)_j.\ee
Let us suppose now that    instead of (2) we use the  decomposition
\be l(\xi^+,\xi^-)=\ti g(\xi^+,\xi^-) h(\xi^+,\xi^-),\ee
where $\ti g\in \ti G$ and $ h\in  G$.
If we assume  that the matrix $E^{ij}(g)$ is invertible
(in the vicinity  of the   origin of the group  $g=e$ this is implied by
the invertibility of $E^{ij}(e)$)
then all the   steps of the previous construction can be repeated.
We end up with the dual $\si$-model
\be \ti L=
\ti E_{ij}(\ti g)
(\ti g^{-1}\partial_+ \ti g)^i(\ti g^{-1}\partial_- \ti g)^j,\ee
where the matrix $\ti E_{ij} (\ti g)$  is defined by ($i=1,\dots,n$)
 \be \ti g^{-1}\ce^+ \ti g ={\rm Span}(\ti T_i + \ti E_{ij}(\ti g) T^j),\ee
\be \ti g^{-1}\ce^- \ti g ={\rm Span}(\ti T_i - \ti E_{ji}(\ti g) T^j),
\ee
and is given by the exact analogue of the formula (9).
 Clearly, at the origin of the group  ($g=e$
and $\ti g =\ti e$ respectively) the matrices $E$ and $\ti E$ are related
as follows
\be E(e)\ti E(\ti e) = \ti E(\ti e) E(e)=1.\ee
This  is an indication that   we have indeed obtained a
 generalization of the
standard Abelian `$R\to 1/R$'  duality.

If the matrix $E^{ij}(e)$ is not invertible, it may seem that the
dual model  does not exist because for the dual decomposition (14) we cannot
find the dual
$\si$-model matrix $\ti E_{ij}(\ti g)$. However, we may proceed as follows.
Let us represent the subspaces $\ti g^{-1}\ce^{\pm} \ti g$ in (16), (17) as
\be \ti g^{-1}\ce^+ \ti g ={\rm Span}(\ti F^{ij}(\ti g)\ti T_j +  T^i),\ee
\be \ti g^{-1}\ce^- \ti g ={\rm Span}(-\ti F^{ji}(\ti g)\ti T_j +  T^i).\ee
The existence of such a matrix $\ti F^{ij}(\ti g)$ is correlated with the
existence of the matrix $E^{ij}(g)$ in (5) or (6).
 In fact,
\be \ti F(\ti e)=E(e).\ee
The explicit dependence of the matrix $\ti F$ on $\ti g$ is again given
by the matrices of the adjoint representation of $D$. Here we have (cf.
(7)--(9))
\begin{eqnarray}\ti g^{-1} \ce^+ \ti g={\rm Span}~
\ti g^{-1}(T^i + \ti F^{ij}(\ti e)\ti T_j)\ti g\cr
={\rm Span}\Big[(\ti F^{ij}(\ti e)\ti a(\ti g)_{j}^{~l} + \ti b(\ti g)^{il})
\ti T_l +\ti d(\ti g)^i_{~l} T^l\Big],\end{eqnarray}
\be \ti g^{-1}\ti T_i \ti g\equiv \ti a(\ti g)_i^{~l} \ti T_l,
\qquad \ti g^{-1} T^i \ti g \equiv
\ti b(\ti g)^{il}\ti T_l +\ti d(\ti g)^j_{~l}  T^l,\ee
\be \ti F(\ti g)=\ti d(\ti g)^{-1}(\ti F(\ti e)\ti a(\ti g)+\ti b(\ti g)).\ee
For the dual
decomposition (14), the field equations which  follow from (1) are
\be \langle \ti g^{-1}\dpm \ti g +\dpm  h  h^{-1},\ti g^{-1}
\ce^{\mp}\ti g\rangle=0.\ee
Using (19),(20)
we obtain
\be (\ti g^{-1}\partial_- \ti g)^i +\ti F^{ij}(\ti g)(\partial_- h h^{-1})_j=0,
\ee
\be (\ti g^{-1}\partial_+ \ti g)^i -\ti F^{ji}(\ti g)(\partial_+ h h^{-1})_j=0.
\ee
This set of equations can be represented  in the following  equivalent form
\be (\ti g^{-1}\partial_- \ti g)^i +\ti F^{ij}(\ti g)\lambda_{-j}=0,\ee
\be (\ti g^{-1}\partial_+ \ti g)^i -\ti F^{ji}(\ti g)\lambda_{+j}=0,\ee
\be  \partial_- \lambda_{+i} -\partial_+ \lambda_{-i} +  c^{kl}_{~~~i}
 \lambda_{+k} \lambda_{-l}=0,\ee
where $c^{kl}_{~~~i}$ are the structure constants of the Lie algebra $\cg$.
In (30) we recognize the zero curvature condition for the `currents'
 in $\cg$.
The equations (28)-(30) are just
the Euler-Lagrange equations corresponding to
\be \ti L=-\lambda_{+i} \ti F^{ij}(\ti g) \lambda_{-j} + \lambda_{+i}
 (\ti g^{-1}\partial_- \ti g)^i +
 \lambda_{-i} (\ti g^{-1} \partial_+ \ti g)^i.\ee
If the matrix $\ti F(\ti g)$ were invertible,  integrating
out $\lambda_\pm$  from (31)  we would get just
 the $\si$-model (15) with
 $$ \ti E(\ti g)=\ti F^{-1}(\ti g). $$
If $\ti F(\ti g)$ does not have the  inverse,
we may integrate out only `non-null' parts of  $\lambda$'s,  while  their
`null-vector'   parts will play
the  role of the  Lagrange multipliers,
constraining  the corresponding projections of the currents $\ti g^{-1}
\partial_{\pm} \ti g$
to vanish. Hence we get a constrained (or `singular')
$\si$-model  defined on the group space $\ti G$.  An example of such  model
will be discussed  in the next section.\footnote{Let us remark that for
$\tilde{F}^{ij}$ being an
 antisymmetric tensor $\tilde{F}^{ij}=-\tilde{F}^{ji}$, the
 $\si$-model (31) becomes topological. The
 consistency of equations of motion then implies the Jacobi
 identity for the tensor $\tilde{F}$ which may be interpreted
 as a Poisson bracket on the target group manifold. Topological
 $\si$-models of this type were introduced as Poisson $\si$-models
 in  \cite{SS,ASS}. In particular, the case of $\tilde{F}^{ij}(e)=0$
 leads to the gauged WZNW model \cite{ASS}. We shall discuss topological
 $\si$-models in the context of Poisson-Lie T-duality elsewhere.}

To summarize,   the Poisson-Lie duality is well defined also
in the case when the $\sigma$-model matrix $E(g)$ of one of the models in the
dual pair  (with the group $G$ as the target) is degenerate.
Then  the dual model  action  can  written in the
first-order form (31) and  can be interpreted as that of the constrained
$\si$-model with the target being the dual group $\ti
G$.\footnote{Alternatively,
by analogy with the gauged WZNW type models containing extra auxiliary vector
fields
(see \cite{Ts2}  and refs. there)   one may trade $\lambda$'s for new fields
and as a result get a model defined on a target of dimension
dim$ \tilde  G$+2$\times$ (number of null-vectors of the matrix $\tilde F$).}
 Similar  situation  happened in the context of the Abelian duality
in a direction of a null isometry.
It thus appears that the Poisson-Lie duality
naturally acts on  the set  of $\si$-models enlarged by the constrained
ones (for a discussion of such models see also \cite{KT,GPR}). That means, in a
sense,   that also $\si$-model matrices
with infinite eigenvalues are to be  included into consideration.

Next, let us comment on  the field equations  for
the pair of  dual models (13) and (15). They both have the form
of the zero  curvature conditions with respect to  the algebras
$\cgt$ and $\cg$.  Such $\si$-models were called
Poisson-Lie symmetric in \cite{KS2,K} and  were shown to be
dualizable with respect to the Poisson-Lie duality. It is
important to be able to express  the corresponding flat  connection
 in terms of the data defining the $\si$-model. As it is clear from
Eq.(30),  the currents are most easily identified
in the first-order formalism (31). Their components are just  the $\lambda$'s
in (31) and this  is true no matter whether the matrix $\ti F(\ti g)$
is regular or degenerate. In the degenerate case, however, we arrive
at an interesting conclusion that some of
the components of the non-Abelian  connection (which is flat according to
equations of motion)
play the role of
the Lagrange multipliers.

It may seem that the Poisson-Lie T-duality relates only $\si$-models
with group targets $G$ and $\ti G$.  However,  it was shown in \cite{KS2,K}
that
this $G \leftrightarrow \ti G$ duality
is only the special case (referred to  as `atomic' duality in  \cite{K})
of the Poisson-Lie
T-duality. When the double is Abelian,  the atomic  duality  is just the
`$R\leftrightarrow 1/R$' duality between the  free  scalar  theories  on
 the group spaces  $U(1)$ and $\tilde U(1)$.
In the Abelian case, however, the notion of T-duality is much broader:
every $\si$-model such that its target is isometric with respect
to the (free) action of the Abelian  duality group,
 is dualizable.  Similar generalization is possible  for the
Poisson-Lie T-duality \cite{KS2,K}:  every $\si$-model such that
a group $G$ acts freely on its target space
and  its action
is Poisson-Lie symmetric with respect to the dual group $\ti G$ has   dual
counterpart
such that $\ti G$ acts freely on its  target space
 and  its  action is Poisson-Lie
symmetric with respect to $G$.
A full classification of the target spaces  admitting
Poisson-Lie symmetry, their description in `adapted' coordinates
(in which the Poisson-Lie symmetry is explicit)  and  the corresponding
 form of the Poisson-Lie
dual target space  are given in \cite{KS2,K}.

\section{\bf $O(2,2)$ double and $SL(2,R)$ WZNW model}
In this section we shall describe an example of a pair of Posson-Lie dual
$\si$-models
which is associated with the $O(2,2)$ double.
Consider the Lie algebra $sl(2,R)$  defined by
\be [H,E_{\pm}]=\pm 2E_{\pm}, \qquad [E_+,E_-]=H, \ee
and equipped with  the standard Killing-Cartan non-degenerate symmetric
invariant
bilinear form
\be \la E_+,E_-\ra=1, \qquad \la H,H\ra=2.\ee
 It can be  checked that the direct sum of the two copies of $sl(2,R)$
 \be \cd =sl(2,R)\oplus sl(2,R)\ee
with the bilinear form (also denoted by $\la .,.\ra$)
\be \la (x_1,x_2),(y_1,y_2)\ra=\la x_1,y_1\ra -\la x_2,y_2\ra\ee
is the algebra of the  Drinfeld double which we shall refer to as the $O(2,2)$
double.
The notation $(x_1,x_2)\in\cd$ obviously means that $x_1$ ($x_2$) is from
the first (second) copy of $sl(2,R)$ in  (34).
The decomposition of the double into the pair of maximally
 isotropic subalgebras
is given as
\be \cd =sl(2,R)_{ diag}  +  b_2\ee
where $sl(2,R)_{ diag}$  is generated by
\be \ti T_0 ={1\over 2}(H,H),\quad \ti T_+= (E_+,E_+), \quad \ti T_-=
 (E_-,E_-)\ee
and $b_2$ (which is the Lie algebra of the Borel subgroup $B_2$ of
$SL(2,C)$)  by
\be T^0={1\over 2}(H,-H), \quad T^+ = (0,-E_-), \quad  T^- =(E_+,0).\ee
These  two sets of generators are dual to each other in the sense of (4).

Starting with the double as the group $D=SL(2,R)\times SL(2,R)$
 we can follow the procedure  of   section 2
and   choose two
mutually orthogonal subspaces $\ce^{\pm}$
\be \ce^+=(sl(2,R),0), \qquad \ce^- =(0,sl(2,R)).\ee
The reason for such a choice is
 that the corresponding field equations (1) are manifestly
chiral,
\be \dpm g_{\mp}=0, \ \ \  \ \ \  l\equiv g_+\times g_-. \ee
Here $g_+$ and $g_-$ are the
 elements from the first and the second copy of
$SL(2,R)$ from the decomposition of the double
and  $\times$ means the direct product of two $SL(2,R)$ elements.
The general solution of (40) is thus
\be l(\xi^+,\xi^-)=g_+(\xi^+)\times g_-(\xi^-)\ . \ee
The appearence  of the  non-Abelian chiral  bosons
suggests a close relation to
WZNW model and thus a possibility of conformal invariance  of  the
 quantum  theory.

To  obtain the dual pair of $\si$-model corresponding to  the choice (39)
we are to  find the explicit form of   Eqs. (5) and (19) in this case.
According to (2), (36)  we may write
\be g_+\times g_- = (b_+ \times b_- )(\zeta\times\zeta)\ee
where $b_+\times b_-\equiv b \in B_2$
and $ \zeta \times \zeta \in SL(2,R)_{diag}.$
 As follows from (38), the ordinary (not direct) product  of the diagonal
parts of the $SL(2,R)$ elements $b_{\pm}$ is the identity element in
$SL(2,R)$.
The corresponding $\si$-model with the Borel group $B_2$ as the target space
is
\be L=E^{ij}J_{+i}(b)J_{-j}(b), \ee
where $ J_{\pm i}(b)T^i$  are defined by
\be b^{-1}\dpm b\equiv J_{\pm i}(b)T^i\ee
and the matrix $E^{ij}$ has the following explicit form
\be E=\left(\matrix{1&0&0\cr 0&0&1 \cr 0&0&0}\right).\ee
 The reason why
the matrix $E^{ij}$ does not
depend on the group  variable $b$  is that
the chosen subspaces $\ce^{\pm}$ are invariant with respect to the adjoint
action on the double.

 Since $E$ is {\it singular},
the dual model should be described by
 the first-order Lagrangian  (31).
Indeed, using  the dual parametrization
of the double (cf. (2),(14),(42))
\be g_+\times g_- = (\eta  \times  \eta)(c_+\times  c_-),\ee
where $\eta \times  \eta \in SL(2,R)_{diag}$  and   $c_+ \times  c_- \in B_2$,
and the relation (21)
we find that (31) takes the form
\be \ti L=-\lambda_{+i} E^{ij} \lambda_{-j} + \lambda_{+i} J_-^i(\eta) +
 \lambda_{-i} J_+^i(\eta).\ee
Here $J_{\pm}^i(\eta)$ are defined by
\be \eta^{-1} \dpm \eta \equiv J_{\pm}(\eta) =
{1\over 2} J_{\pm}^0(\eta) H +
J_{\pm}^+(\eta) E_+
+J_{\pm}^-(\eta) E_-.\ee
Let us  now look more closely at  the pair of the mutually dual models
(43) and (47).
The first Lagrangian  (43) can be rewritten in terms of the $sl(2,R)$-currents
$J_{\pm}^i(b_{\pm})$  defined as in (48)
\be L(b)=-J_+^0(b_+)J_-^0(b_-)-J_+^-(b_-)J_-^+(b_+).\ee
This   follows from the obvious relations
\be J_0(b)=J^0(b_+)=-J^0(b_-), \ \  J_+(b)= -J^-(b_-), \ \  J_-(b)=
J^+(b_+),\ee
where $J_i(b)$ were  defined in (44) (we suppress the 2d space-time indices of
these currents).

The action corresponding to  (49) is nothing but
the $SL(2,R)$ WZNW action $I$  for  the argument
 $b_- b_+^{-1}$ (we shall not explicitly indicate the standard
measure of integration $d\xi^+ d\xi^-$)
\be \int  L(b)= -4I(b_- b_+^{-1}), \ee
where
\be I(u) \equiv {1\over 8\pi }\int_{\partial M} {\rm Tr}~(\partial_+ u
\partial_- u^{-1})
+{1\over 12\pi }\int_M  {\rm Tr}~(u^{-1}du)^{\wedge 3} .\ee
One can easily check this  using the  Polyakov-Wiegmann relation  \cite{Pw}
\be I(ff')=I(f) +I(f') -{1\over 4\pi }\int {\rm Tr}(f^{-1}
\partial_+ f \partial_- f' f'^{-1}).\ee
The conclusion is that  the first  (43) in  the  pair of dual models is
actually
the WZNW model on the group
manifold $SL(2,R)$. Indeed,
the combination $b_- b_+^{-1}$
can be interpreted as the Gauss decomposition of a group  element $u$
 parametrizing the $SL(2,R)$ group space.
 The  Gauss decomposition $u
=b_-b_+^{-1}$  was used  \cite{GAB}
in representing the  WZNW  theory in terms of free fields. We have gone in  the
opposite direction, starting from the simple action (49) on the Borel group
$B_2$ and recovering the WZNW action  at the end.

The fact that the  $\si$-model on the Borel group  has an
interpretation in terms of the $SL(2,R)$ WZNW model is  quite interesting.
 By construction,  the dual model  also  has the  $SL(2,R)$ group
as a target space. Thus the duality relates two
different $\si$-models  (unconstrained and constrained one) defined on the
$SL(2,R)$ space.
Under an appropriate choice of the currents, the    $\si$-model matrix of one
model
 is  the inverse (in a loose sense, since $E$ is degenerate)
of the matrix of the other. This is true globally, i.e. not only at the group's
origin as in the generic Poisson-Lie duality case (18).

The  Lagrangian
 (47) of the second model can be put in  a simpler form by integrating out all
$\lambda$'s
except the Lagrange multipliers (those which drop out from the first term in
(47)). The result is  the Lagrangian of the constrained $\si$-model\footnote{As
 remarked
at the end of section 2, the Lagrange multipliers are  the components of the
(flat) connection so it is convenient to write them also with the Lie
algebra indices.}
\be \ti L(\eta)= J_+^0(\eta)J_-^0(\eta) + J_-^+(\eta)J_+^-(\eta)+
\lambda_-^+ J_+^+(\eta)+\lambda_+^- J_-^-(\eta). \ee
We shall give the explicit form of (54)  (in a particular parametrisation of
$\eta$)  in the next section.

\section{\bf Duality and path integral}

So far our discussion was purely classical -- we have demonstrated
the duality between the pair of the $\si$-models with the targets $G$ and
$\ti G$ at the level of equations of motions. It would be highly desirable
to find a path integral formulation of the Poisson-Lie T-duality and
 establish a quantum equivalence of the models.
This, indeed, is  possible to do in the example discussed in the previous
section.

Consider the following action  \cite{Ts,Ts2}
\be S(g_+,g_-) =  k[I_+(g_+) + I_-(g_-)], \ee
where $k$=const and $I_{\pm}(g_\pm) $ are defined by ($\partial_{0,1} \equiv
{1\over 2} (\partial_+ \pm \partial_-)$)
\be I_{\pm}(g_{\pm}) =\pm {1\over 4\pi }\int_{\partial M} {\rm Tr}~
(\partial_1 g_{\pm}^{-1}
 \partial_{\mp} g_{\pm})
+{1\over 12\pi }\int_M  {\rm Tr}~(g_{\pm}^{-1}dg_{\pm})^{\wedge 3} , \ee
 or, equivalently, by ($I$ is  the WZNW action (52))
\be I_-(g_-)=I(g_-)-{1\over 8\pi }\int {\rm Tr}
(\partial_+ g_- \partial_+ g_-^{-1}),\ee
\be I_+(g_+)=I(g_+)-{1\over 8\pi }\int {\rm Tr}
(\partial_- g_+ \partial_- g_+^{-1}).\ee
The actions $I_{\pm}$ \cite{Son,Ts}  describe
the non-Abelian chiral scalars. The corresponding   field equations
$\partial_1(g_{\mp}^{-1}\dpm g_{\mp})=0$
imply
\be \dpm g_{\mp}=0,\ee
provided the fields $g_{\pm}(\xi^+,\xi^-)$
are  subject to appropriate boundary conditions \cite{FJJ}. The latter should
be such that the equation $\partial_1 f=0$
should have  the unique solution  $ f=0,$ i.e.
\be \partial_1 f(g_+,g_-)=0  \ \ \rightarrow \ \ f(g_+,g_-)=0.
\ee
The actions (55)--(58)
are of first-order type, i.e. are linear in the time derivatives of the fields.
 It is therefore natural  to  try to integrate one  `half'
of the field variables in (55)
to end up with a second-order and Lorentz invariant
action.  In fact, starting with (55) it is possible to integrate out the
`ratio'
of $g_-$ and $g_+$ explicitly, ending up with the standard WZNW action for
$g_-g_+$ \cite{Ts}.

As  we  shall see below,
there are two dual choices of
what should be the `half' of the variables to be integrated away.
In the case of the double $O(2,2)$ these
choices  lead precisely to the dual pair of actions (43) and (47)
considered in
the previous section.

The equations (59) are identical to
 the basic equations  (40) on the double. Thus  we may attempt
to use the action $S(g_+^{-1} ,g_-)$ on the double  as  the one which
 `interpolates'
between the  two dual $\si$-models.
This basic idea can be illustrated on the
 simplest  $U(1) \times U(1)$ free-theory case ($g_\pm= e^{ix_\pm}$) \cite{Ts}
\be S=k[ I_+(x_+) + I_-(x_-)],  \ \ \
I_{\pm}(x_{\pm}) =\pm {1\over 4\pi }\int
\partial_1 x_{\pm}
 \partial_{\mp} x_{\pm}.
\ee
Introducing the new fields $x,\tilde x$
\be x={1\over \sqrt{2}R}(x_+ +x_-), \qquad \ti x={R\over \sqrt{2}} (x_+-x_-)\ee
we get the action
\be S= {k\over 4\pi} \int (\partial_0 x \partial_1 \tilde x
+ \partial_1 x \partial_0 \tilde x  - R^2 \partial_1 x \partial_1  x
-  R^{-2} \partial_1 \tilde  x \partial_1 \tilde x)  , \ee
which is invariant under $x \to \tilde x, \tilde x \to x$, $ R\to 1/R$.
Integrating away  in the path integral  the variable $x$ or $\ti x$,
one  obtains  the   pair of free scalar actions related by the standard
Abelian duality.

We shall argue that the non-Abelian generalizations of the relations  (63)
are  given by the products (2) and (14)  in the Drinfeld double. Note
that the treatment of  the non-Abelian case  requires the  introduction of the
four variables
$g,\ti g$ and $h,\ti h$
as opposed to  the two variables $x$ and $\ti x$ in (63). The reason  for that
is that
 $g$ ($\ti g$) does not  commute with $\ti h$ ($h$).

Consider now
 the path integral for the non-Abelian action $S(g_+^{-1},g_-)$ (55)
\be Z=\int [dg_+][dg_-] \exp{[iS(g_+^{-1},g_-)]}\ee
and assume that the  first  parametrization (42) is used.
Using the Polyakov-Wiegmann relation  (53)
 we  get
\be Z =\int [db][d\zeta] \exp{ik[I(b_- b_+^{-1}) - {1\over 8\pi}
\int {\rm Tr}(2\partial_1 \zeta \zeta^{-1} - J_-(b_+) + J_+(b_-))^2]}.
\ee
The integration over $\zeta$ gives a  trivial
contribution \cite{Ts} since one can replace the integral over $\zeta$
by the integral over
$B=\partial_1 \zeta \zeta^{-1}$ (the resulting Jacobian  is equal to one
under  the choice of the boundary conditions (60)).\footnote{The
determinant of the operator $\partial_1- B$
should be computed using the  Green fuction $\theta(x_1-x'_1)$
of $\partial_1$.  This choice makes the problem essentially a one-dimensional
one and ensures the absence of anomaly.}
 As a result,
\be Z= \int [db] \exp{[ikI(b_- b_+^{-1})]},\ee
i.e.  we  get precisely the first model (51) of our dual pair.

To obtain  the  dual model we have to start with the parametrization
(46) and integrate away the fields $c_{\pm}$.
Choosing  the  parametrization of $c_{\pm}$ such that
\be g_-=\eta c_- =\eta \left(\matrix{e^{\chi}&0\cr 0& e^{-\chi}}\right)
\left(\matrix{1&0\cr \theta &1}\right)\equiv \eta_-
\left(\matrix{1&0\cr \theta &1}\right),\ee
\be g_+=\eta c_+ = \eta \left(\matrix{e^{-\chi}&0\cr 0& e^{\chi}}\right)
\left(\matrix{1&\rho\cr 0 &1}\right)\equiv \eta_+
\left(\matrix{1&\rho\cr 0 &1}\right), \ee
we get
\be I_-(g_-)=I_-(\eta_-) +{1\over 2\pi }\int J_+^+(\eta_-)\partial_1\theta,\ee
\be I_+(g_+^{-1})=I_+(\eta_+^{-1}) -{1\over 2\pi }
\int J_-^-(\eta_+)\partial_1\rho,\ee
where the current components  were defined  in (48).
Thus
\begin{eqnarray} Z=\int [d\eta][d\chi][d\theta][d\rho]\exp{ik[I_-(\eta_-)+
I_+(\eta_+^{-1})]}\cr \times \exp{\big({1\over 2\pi }ik \int \big[
J_+^+(\eta_-)
\partial_1\theta
-  J_-^-(\eta_+)\partial_1\rho\big]\big). }\end{eqnarray}
Using that
\be I_-(\eta_-)=I_-(\eta)+{1\over 2\pi }\int ( \partial_+\chi \partial_1\chi
+ J_+^0(\eta) \partial_1\chi),\ee
\be I_+(\eta_+^{-1})=I_+(\eta^{-1}) -{1\over 2\pi }\int( \partial_-\chi
\partial_1\chi-  J_-^0(\eta)\partial_1\chi), \ee
and  the Polyakov-Wiegmann relation  (53),
the total action can be  rewritten as
\be S(g_+^{-1},g_-)  ={1\over 4\pi }k
\int \Big[- (J_+^0(\eta)J_-^0(\eta)+J_+^-(\eta)J_-^+(\eta)
+J_+^+(\eta)J_-^-(\eta))\ee $$
+ \  {1\over 4} (4 \partial_1\chi+
 J_+^0(\eta)+ J_-^0(\eta))^2   $$  $$
+ \  ( J_+^-(\eta)+2 e^{-2\chi}\partial_1 \theta)J_+^+(\eta)
+  (J_-^+(\eta) -2 e^{-2\chi}\partial_1 \rho)J_-^-(\eta)\Big].  $$
We can now
 perform the  change of variables in the path integral from $\theta$
and $\rho$ to $\lambda_-^+$ and $\lambda_+^-$ defined by
\be \lambda_-^+ = - J_+^-(\eta)-2 e^{-2\chi}\partial_1 \theta,\ee
\be \lambda_+^- = -J_-^+(\eta) +2 e^{-2\chi}\partial_1 \rho.\ee
The corresponding  Jacobian is trivial  (under the assumption of the boundary
conditions (60) as in (65),(66),  implying the  Lorentz invariance of the total
path integral).
Then  we  are able to  integrate away
the $\chi$-dependence.  After a  shift of $\lambda_+^-$
 we  finally obtain
\be Z = \int [d\eta][d\lambda^-_+][d\lambda^+_-]
\exp{\Big( - {1\over 4\pi}ik\int [
 J_+^0(\eta)J_-^0(\eta) }\ee $$   + \  J_-^+(\eta)J_+^-(\eta)+
\lambda_-^+ J_+^+(\eta)+\lambda_+^- J_-^-(\eta)]\Big),  $$
which is precisely the path integral corresponding to  our dual model
(54).\footnote{Note that the  overall
constant $k$ was not changed to $1/k$.
 This does  not happen  also  in the case of the
Abelian T-duality (cf. (62) and (63)).
It is true  that the Abelian duality (and in general
Poisson-Lie duality, see (18)) change  `part' of the coupling matrix
into its  inverse,  but not the { overall} coefficient (we consider the `local'
duality transformations, i.e. ignore the issue of zero modes).}

The original action $S$ (55) is conformal,
being the sum of two chiral WZNW actions (in particular, $Z$ in (64) does not
depend on the conformal factor of the 2-metric apart from the overall
 `central charge' term).
This means that the actions we got  from (64)  by partial integrating out
the subsets of variables are also conformal.

The action (54) and the corresponding path integral (77) can be put in more
explicit form by using the
following parametrization of $\eta\in SL(2,R)$
\be \eta=
\left(\matrix{1&0\cr u&1}\right)\left(\matrix{1&v^{-1}\cr 0&1}\right)
\left(\matrix{e^{\phi}&0\cr 0&e^{-\phi}}\right),\ee
\be J_-^-(\eta)=e^{2\phi}\partial_-u,\qquad J_+^+(\eta)=-e^{-2\phi}v^{-2}
\partial_+ (u+{v}).\ee
The  classical  constraints  $J_-^- =0, \ J_+^+=0$  then  take the
chiral form
\be \partial_-u =0,\qquad \partial_+ (u+{ v})=0, \qquad u= u_+ (\xi^+), \ \
v=v_-(\xi^-) - u_+(\xi^+).\ee
Under these constraints, the  remaining   components  of the currents are
\be J_-^+=e^{-2\phi}\partial_- v^{-1} ,\ \  J_+^- =-e^{2\phi}\partial_+ { v}, \
\
  J_-^0=2\partial_-\phi, \ \  J_+^0 = 2\partial_+ (\phi +\ln v), \ee
so that the  Lagrangian  in (54)  becomes
\be  L'=  J_+^0 J_-^0   +   J_-^+ J_+^-  =    4\partial_+ (\phi  + \ln v)\
\partial_- \phi
 +\partial_+ \ln v \ \partial_-\ln v, \ee
and thus describes one off-shell scalar degree of freedom $\phi$ coupled to
$u_+$ and $v_-$.

After $\lambda$'s and $u,v$ are integrated over in the path integral
(77),  one gets also the  contributions of the determinants of the operators
$e^{\pm 2\phi}\partial_\mp$  which lead (as in the last reference in
\cite{GAB})
to the shift  of the  coefficient of the $\partial_+ \phi\partial_- \phi$
term and to the  2d curvature coupling $R^{(2)} \phi$
 which combine to  reproduce the  central charge of the $SL(2,R)$ WZNW
model.\footnote{Let us note that the Lagrangian of the $SL(2,R)$ WZNW model
in the standard Gauss decomposition parametrisation takes the form
$L= k( \partial_+ \psi \partial_- \psi  +
 e^{-2\psi} \partial_+ u' \partial_- v'), $
so that integrating out $u'$ and $v'$
one gets the free action for $\psi$ with shifted $k'=k-2$
and  linear 2d curvature coupling term (see Gerasimov et al, ref. \cite{GAB}).
Similar `reduction' is possible for the chiral actions $I_\pm$ in (56).
One finds that similar   effective actions for the Cartan variables
have the structure $ \int[(k-2)  \pm \partial_1 \psi_\pm \partial_\mp \psi_\pm
+ \sqrt {g} R^{(2)} \psi_\pm]$. Starting from the sum of these actions  (55)
and integrating out the combination
 $\psi_+ -\psi_-$
one finds the  corresponding  action for  the Cartan part  $\psi = {1\over
\sqrt 2} (\psi_+ +\psi_-)$ of the WZNW model,
cf. (61)--(66). }

It should be noted that this reduction
does not imply, however,
 that the  $SL(2,R)$ WZNW model is equivalent to
 that  of  the  free $\phi$-theory.
 Indeed, (77) is just the  vacuum partition
function,
 while to have a  complete  picture of duality at the quantum level
one is to compare  the generating functionals
for  appropriate correlation functions.
The correlators may contain the fields  which  are
simply integrated out in  the  vacuum case (in particular,  they  are likely
to depend on
the `Lagrange multipliers' $\lambda$'s  in (54),(77)).

\section{\bf Outlook}

It is very likely  that the method we have  described above
can be used for the construction of Poisson-Lie T-dual of any
maximally noncompact WZNW model (see \cite{KT} and refs. there).
Another problem is to  find a dictionary
between  correlators of local operators in the two dual  models. This  would
 reveal a nontrivial  quantum content of the  duality symmetry.
In order to complete the full analogy between the standard Abelian
and Poisson-Lie duality it would be also important to understand
the issue of the zero modes which was not  addressed  here.

\section{\bf Acknowledgements}
We thank K. Gaw\c edzki, E. Kiritsis, P. \v Severa,
and S. Shatashvili
 for useful discussions. A.A.T. acknowledges the hospitality of CERN
Theory Division and the support of PPARC, ECC grant SC1*-CT92-0789 and
NATO grant CRG 940870.

\end{document}